\pdfoutput=1
\documentclass[12pt,aps,longbibliography,superscriptaddress]{revtex4-1}
\usepackage[colorlinks=true,citecolor=blue,urlcolor=blue,linkcolor=blue,pdfstartview=FitH,bookmarksopen]{hyperref}
\usepackage{amsmath,amssymb}
\usepackage{bm}
\usepackage{graphicx}
\usepackage{xcolor}
\usepackage{comment}
\usepackage{caption}
\usepackage{subcaption}

\newcommand{\be}{\begin{equation}}
\newcommand{\ee}{\end{equation}}
\newcommand{\bear}{\begin{eqnarray}}
\newcommand{\eear}{\end{eqnarray}}
\newcommand{\ba}{\begin{array}}
\newcommand{\ea}{\end{array}}
\renewcommand\d{\partial}

\DeclareMathOperator\Li{Li}
\newcommand\+\dagger
\newcommand{\m}{\widetilde m}

\begin{document}
\title{The phase diagram of ultra quantum liquids}
\author{Dam Thanh Son}
\affiliation{Kadanoff Center for Theoretical Physics, University of Chicago, Chicago, IL 60637, USA}
\author{Mikhail Stephanov}
\affiliation{Department of Physics, University of Illinois, Chicago, IL 60607, USA}
\author{Ho-Ung Yee}
\affiliation{Department of Physics, University of Illinois, Chicago, IL 60607, USA}
\affiliation{Kadanoff Center for Theoretical Physics, University of Chicago, Chicago, IL 60637, USA}

\begin{abstract}
We discuss the dependence of the phase diagram of a hypothetical
isotope of helium with nuclear mass less than $4$ atomic mass units.
We argue that with decreasing nucleus mass, the temperature of the
superfluid phase transition (about 2.2 K in real $^4$He) increases,
while that of the liquid-gas critical point (about 5.2 K in real
$^4$He) decreases.  We discuss various scenarios that may occur when
the two temperatures approach each other and the order parameters of
the superfluid and the liquid-gas phase transitions interact with each
other.  The simplest scenario, in which both order parameters become
critical at particular values of the nuclear mass, temperature, and
pressure, can be ruled out through on an analysis of the Landau theory.
We argue that in the most likely scenario, as the nuclear mass
decreases, first, a tricritical point appears on the line separating
the superfluid and the normal fluid phase, then the critical point
disappears under the first-order part of superfluid phase transition
line, and in the end the tricritical point disappears.  The last
change in the phase diagram occurs when the two-body scattering length
crosses zero, which corresponds to the nuclear mass of about 1.55~u.
We develop a quantitative theory that allows one to determine the
phase diagram in the vicinity of this point.
Finally, we discuss
several ways to physically realize such liquids.

\end{abstract}
\maketitle

\section{Introduction}

Helium is a prototypical quantum liquid, being the only natural
substance that remains liquid down to zero
temperature~\cite{Khalatnikov}.  The reason for this behavior is the
weakness of the interatomic potential and the smallness of the atomic
mass, leading to large zero-point fluctuations destroying the would-be
crystal.

\begin{figure}[h]
\begin{center}
\includegraphics[width=6cm]{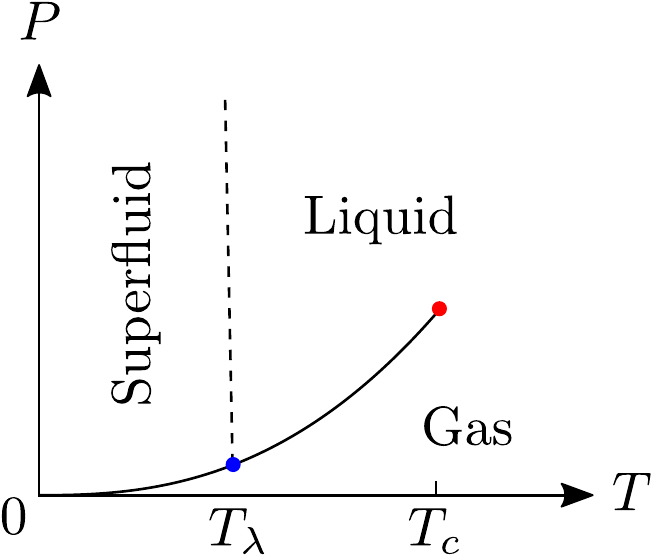}
\caption{The schematic phase diagram of physical $^4$He.  The solid phase at
  higher pressure is outside of the frame.  Solid lines denote first-order
  phase transitions, dashed lines denote second-order phase
  transitions.
}
\label{fig:physical_pd}
\end{center}
\end{figure}

The phase diagram of $^4$He is schematically depicted in
Fig.~\ref{fig:physical_pd}.  The solid phase at high pressure has been
left out of the phase diagram; we focus on the
gas, the normal fluid, and the superfluid phases.  There are two
characteristic temperatures that can be seen from this phase diagram.
The first is the temperature of the superfluid-to-normal phase
transition,~$T_\lambda$.  This phase transition is second order and
occurs at roughly $T\approx 2.2$ K (the temperature depends slightly
on the pressure).  The second temperature is the temperature of the
liquid-gas critical point, $T_c$ which is at approximately
5.2~K.

One can ask whether this phase diagram should be expected for any
fluid which becomes a superfluid at low temperature.  (Here by
``fluid'' we mean a thermodynamic system that has a liquid-gas phase transition
terminated at a critical point.
Typically, such a system is made out of particles which
interact which each other through a potential with a van-der-Waals-type
long-distance attractive tail and a short-range repulsive core.)
Because helium is a unique liquid
that becomes a superfluid at low temperature, this question is rarely asked.
But one can, for example, inquire if the liquid-gas critical
point can be made to lie inside the superfluid phase.

In this paper we try to establish the possible phase diagrams of an ``ultra-quantum
liquid,'' which can be thought of as a hypothetical isotope of helium
with a bosonic nucleus lighter than the $^4$He nucleus (the alpha particle).
Assuming that the mass of the nucleus is still much
larger than the mass of the electron, one can use the Born-Oppenheimer
approximation to treat the motion of the nuclei.  In particular, the
interaction potential between the atoms remains unchanged.
For the purpose of our discussion, which is focused on the qualitative
features of the phase diagram, we can replace the exact interaction
potential between the helium atoms by the Lennard-Jones potential
\begin{equation}\label{LJpotential}
  V(r) = 4\epsilon \left[ \left( \frac\sigma r\right)^{12}
    - \left( \frac\sigma r\right)^6 \right],
\end{equation}
which reaches a minimum of $-\epsilon$ at $r=2^{1/6}\sigma$.  We expect
that the precise form of the potential should not matter much for
qualitative questions; the potential~(\ref{LJpotential}) can be
replaced by any potential with a $r^{-6}$ long-distance tail and a
short-distance repulsive core.

  The
magnitude of quantum effects is parametrized by the de Boer parameter~\cite{deBoer1948}
\begin{equation}
  \Lambda = \frac\hbar{\sigma\sqrt{{m}\epsilon}} \,.
\end{equation}
In the WKB approximation, $\Lambda$ is inversely proportional
to the number of bound states that the potential supports, so
a larger $\Lambda$ corresponds to a more quantum system.
The standard values of the Lennard-Jones parameters for helium are~\cite{BerryRiceRoss}
\begin{equation}\label{LJ-numerical}
  \sigma=2.556~\text{\AA}, \qquad \epsilon/k_B=10.22~\text{K}\,.
\end{equation}
For the physical mass $m=m_{\rm phys}=4$~u, $\Lambda\approx 0.426$.
(For comparison, hydrogen H$_2$ has the de Boer parameter of
about 0.275 and is thus less quantum that $^4$He.)

Often one defines the van der Waals length through the coefficient of the  
$r^{-6}$ tail of the potential, $V(r)\sim -C_6 r^{-6}$ as $r\to\infty$, as
\begin{equation}
  \ell_{\rm vdW} = \frac 12 \left( \frac{mC_6}{\hbar^2} \right)^{1/4} .
\end{equation}
A particle rolling down the $C_6r^{-6}$ potential from infinity with
zero energy acquires a WKB phase of order unity when it reaches the radius
$\ell_{\rm vdW}$.
The de Boer parameter can be  written as
\begin{equation}
  \Lambda = \frac{\sigma^2}{2\ell_{\rm vdW}^2} \,.
\end{equation}

As $m$ decreases, $\Lambda$ increases.  Now, since the quantum effects
are more enhanced than in $^4$He, we expect than the ground state at
zero temperature and zero pressure is not a solid for all $m<m_{\rm
  phys}$.  We want to understand, at the qualitative level at least,
what happens to the phase diagram as the fluid becomes more quantum.
(Previously, the effect of changing the nuclear mass on the
zero-temperature ground state of helium has been studied in
Refs.~\cite{Nosanow1975,Miller1977,Egger2011}.)

Qualitatively, one may speculate that the hierarchy $T_\lambda <
T_c$ may no longer hold as $m$ decreases.  Very roughly, the
superfluid phase transition can be interpreted as the Bose-Einstein
condensation, which occurs at temperature
\begin{equation}
  T_{\rm BEC} = \frac{\hbar^2 n^{2/3}}m\,.
\end{equation}
Assuming that the density $n$ of the fluid does not change much (which
is a reasonable assumption since the distance between atoms in the
liquid phase is determined by the location of the minimum of the
interatomic potential $\sigma$), with decreasing $m$ then the
superfluid transition occurs at higher and higher temperature.

On the other hand, for a classical gas the $m$ dependence of the
thermal partition function can be completely factored out, so the
location of the critical point should be independent of the nuclear
mass.  But when the critical temperature is not much larger than the
temperature of the superfluid phase transition, one needs to take into
account quantum effects on the position of the critical point.
Studies of the isotopic dependence of the critical point show that the
critical temperature decreases with decreasing mass (or increasing de
Boer parameter)~\cite{Young:1980}.  So at some value of the mass $m$ the
superfluid transition temperature may become equal or even larger than
than the liquid-gas critical temperature. How does the phase diagram
looks like then?

\begin{figure}
\begin{center}
\includegraphics[width=6cm]{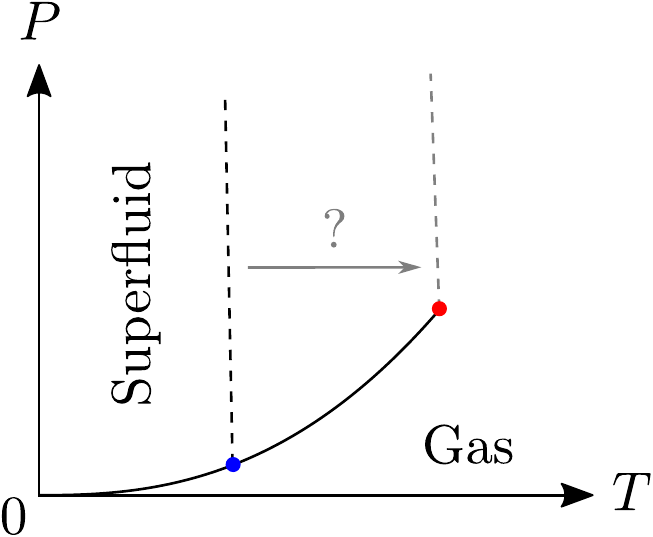}
\caption{The hypothetical simplest scenario. We show that it is not realized.}
\label{fig:hypothetical_pd}
\end{center}
\end{figure}

The simplest scenario is depicted in Fig.~\ref{fig:hypothetical_pd}.
As $m$ decreases the distance between the critical end point (the
point and the critical point becomes smaller, and vanishes at some
value of $m$.  At this critical value of $m$ then, both the superfluid
and the liquid-gas order parameters would become critical at the same
point on the $(P,T)$ phase diagram.

\begin{figure}
  \centering
  \begin{subfigure}{.33\textwidth}
    \centering
    \includegraphics[width=4.75cm]{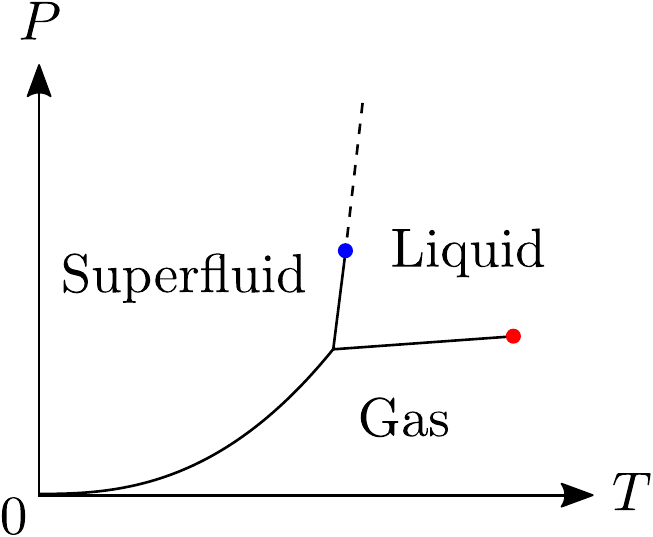}
    \caption{}
    \label{fig:phasediagram1}
  \end{subfigure}%
  \begin{subfigure}{.34\textwidth}
    \centering
    \includegraphics[width=4.75cm]{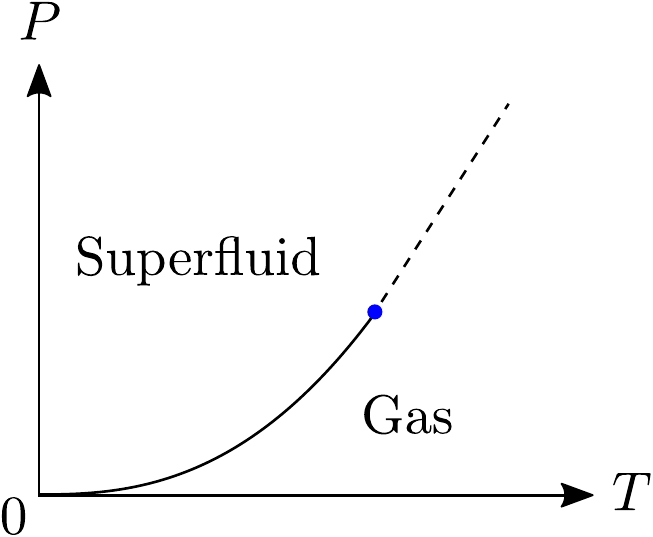}
    \caption{}
    \label{fig:phasediagram2}
  \end{subfigure}%
  \begin{subfigure}{.33\textwidth}
    \centering
    \includegraphics[width=4.75cm]{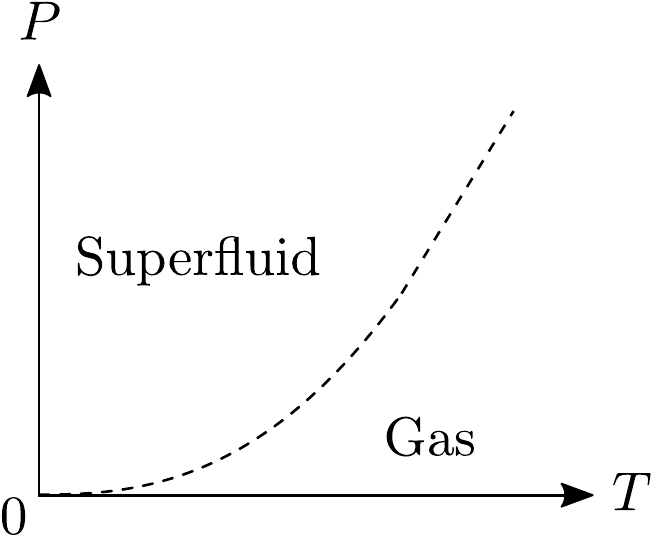}
    \caption{}
    \label{fig:phasediagram3}
  \end{subfigure}
\caption{The evolution of the phase diagram as one lowers the atomic mass.}
\label{fig:3diagram}
\end{figure}

In this paper we rule out this simplest scenario.  We argue that as
the nuclear mass is lowered, the following succession of changes in the
topology of the phase diagram occurs.  The boundary between different
regimes are denotes as $m_1$, $m_2$, $m_3$.
\begin{itemize}
\item For $m_1 < m < m_{\rm phys}$ the phase diagram is topologically
  the same as that of the physical $^4$He (Fig.~\ref{fig:physical_pd}).
\item For $m_2<m<m_1$ a tricritical point appears on the line of the
  superfluid phase transition.  The superfluid phase transition,
  instead of being always second-order, is now first-order at low
  pressure and second order at high pressure.  The junction between the
  superfluid, normal fluid and gas phases is now a triple point instead
  of being a critical endpoint~\cite{Fisher-1991} (Fig.~\ref{fig:phasediagram1}).
\item When $m_3 < m <m_2$, the liquid gas critical point disappears
  under the first-order superfluid phase transition line.  The phase
  diagram now has two phases, a superfluid phase and a normal phase,
  separated by the line of phase transitions which are first-order at low
  pressure and second-order at high pressure (Fig.~\ref{fig:phasediagram2}).
\item For $m< m_3$ the tricritical point disappears, and the whole
line of superfluid phase transition is second order (Fig.~\ref{fig:phasediagram3}).
\end{itemize}

While $m_1$ and $m_2$ cannot be computed analytically, the value of
$m_3$ can be determined: it is the value of the mass for which the
scattering length for the two-atom scattering vanishes.  For
helium this happens when $m$
is $1.55$ atomic units.  Thus, all the transformations of
the phase diagram described above happen in a relatively narrow window
of nuclear mass, ranging from $1.55$ to 4 atomic units.

\section{Analysis of the mean-field theory}

\subsection{Ruling out the simplest scenario}

We now show that the simplest scenario, in which at a critical value
of the mass the phase diagram looks like in
Fig.~\ref{fig:hypothetical_pd}, is excluded.  In this scenario one
would have a
multicritical point with $\text{O}(2)\times\text{Z}_2$ symmetry.  We first analyze
this hypothetical multicritical point from the point of view of Landau's
theory of phase transition.

We introduce two order parameters: the superfluid condensate $\psi$
and the liquid-gas order parameter $\phi$.  The condensate $\psi$ is complex and the
free energy is supposed to have U(1) symmetry $\psi\to\psi
e^{i\alpha}$.  On the other hand, there is no symmetry associated with
the order parameter $\phi$ of the liquid-gas phase transition, which
can be taken to be the density.  Writing down all terms in the free
energy to forth order in the expansion in powers of $\psi$ and $\phi$, we get
the most general expression
\begin{equation}\label{Omega-beta}
  \Omega(\psi,\phi) = \frac t2 \phi^2 + \frac u4 \phi^4 - h\phi
  + (t+\m)|\psi|^2 + \frac\lambda2 |\psi|^4 
  -\alpha\phi|\psi|^2
  -\gamma \phi^2|\psi|^2.
\end{equation}
Here we have eliminated the (redundant) $\phi^3$ term by using the freedom to
shift $\phi$ by an arbitrary constant $\phi\to\phi+c$ (note that there
is no symmetry $\phi\to-\phi$).  

The $\text{O}(2)\times\text{Z}_2$ multicritical point is achieved when the coefficients
of $t=\m=h=\alpha=0$.  This requires fine tuning of four variables,
while we can change only three parameters (the temperature, the pressure, and the mass of
the nucleus).  Therefore, the $\text{O}(2)\times\text{Z}_2$ multicritical point cannot
be achieved.


The argument given above is essentially based on mean-field theory
neglecting critical fluctuations.  To show that our
conclusion remains valid beyond the mean-field level, one needs to
show that each of the six fixed points~\cite{Nelson1974} of the model
has {\em more than three\/}
O(2)-symmetric (but not necessarily Z$_2$ symmetric) relevant
operators.   Let us do that first for the O(3) fixed point, where
an enhanced symmetry combines
the superfluid order parameter $\psi$ and liquid-gas order
parameter (appropriately shifted and rescaled) $\chi$ into a
O(3) vector: $\Phi_a\sim (\text{Re}\,\psi, \text{Im}\,\psi, \phi)$.
The O(3) fixed point has the
following relevant operators, ordered by the rank of the representation
of the SO(3) group~\cite{Hasenbusch2011},
\begin{equation}
\begin{split}
    & \Phi^2 \equiv \Phi_a\Phi_a;\\
    & \Phi_a ;\\
    & \Phi_{ab} = \Phi_a\Phi_b - \frac13 \delta_{ab}\Phi^2; \\
    & \Phi_{abc} = \Phi_a\Phi_b\Phi_c - \frac15 \Phi^2 
     (\delta_{bc}\Phi_a + \delta_{ca} \Phi_b + \delta_{ab}\Phi_c)\,.
\end{split}  
\end{equation}
All these operators have dimensions below 3; for example, the
dimension of the operator $\Phi_{abc}$ has been estimated to be close
to 2~\cite{Hasenbusch2011}.  (In addition, the rank-4 operator
$\Phi_{abcd}=\Phi_a\Phi_b\Phi_c\Phi_d+\cdots$ may be very weakly
relevant~\cite{Calabrese2003}).  
Restricting only to O(2) symmetric-operators, we have the
following relevant operators at the O(3) fixed point
\begin{equation}\label{O2-operators}
\begin{split}  
   \Phi^2 & \sim |\psi|^2 + \phi^2;\\
    \Phi_3 & \sim \phi ;\\
    \Phi_{33} &\sim 2 \phi^2 - |\psi|^2;\\
    \Phi_{333} &\sim 2 \phi^3 - 3 |\psi|^2 \phi\,.
\end{split}    
\end{equation}
(There would be one more relevant operator if the rank-4 operator is relevant
at the O(3)-invariant fixed point.)  Thus there are at least four
relevant operators at the O(3) symmetric fixed point.

The other coupled fixed point is the biconical fixed point~\cite{Nelson1974}.
This fixed point was found to be very close to the O(3) critical
point~\cite{Calabrese2003}, so the operators (\ref{O2-operators})
should be relevant also at the biconical fixed point.  Thus one
also needs {\em at least four\/} fine-tuned parameters to reach this fixed point.  The
analysis can be performed easily at the other four fixed points, where
the order parameters decouple and the operator dimensions are just
sums of operators dimensions in the Z$_2$ and O(2) theories.
One concludes
that there is no $\text{O}(2)\times\text{Z}_2$ fixed point that
can be approached by fine-tuning three O(2) symmetric parameters.

\subsection{The overall evolution of the phase diagram}

To understand what actually happens as one lowers the nuclear mass, we analyze the mean-field theory
based on the free energy~(\ref{Omega-beta}).  For simplicity we set $\gamma=0$ and
rescale the fields so that $\alpha=1$.  The thermodynamic potential in a
grand canonical ensemble (the negative of pressure) that needs to be
minimized in the phase diagram in $(t,h)$ space is
\be\label{Omega}
\Omega(\phi,|\psi|)={t\over 2}\phi^2+{u\over 4}\phi^4-h\phi+(t+\m)|\psi|^2+{\lambda\over 2} |\psi|^4-\phi|\psi|^2.
\ee
Here, $(u,\lambda)$ are assumed to be positive constants, but $\m$ is
a variable parameter that can have any sign, and is presumably related
to the mass of a helium atom. The $(t,h)$ should be related to the
conventional parameters $(T,\mu)$ in the ordinary phase diagram.

We now investigate numerically the evolution of the phase diagram of the
model given by Eq.~(\ref{Omega}) as a function of $\m$.
To find the minimum of the free energy, one first notices that
$\Omega$ is quadratic in $|\psi|^2$, so it is easy to ``integrate
out'' $\psi$, which results in the replacement in Eq.~(\ref{Omega}) of
$|\psi|^2$ by 
its value at the minimum given by
\be
|\psi|^2={1\over\lambda}(\phi-\m-t)\theta(\phi-\m-t)\,,\label{psi2}
\ee
where $\theta(x)$ is the Heaviside step function. 
Then the resulting effective potential for $\phi$ becomes
\be
V(\phi)={t\over 2}\phi^2+{u\over 4}\phi^4-h\phi-{1\over 2\lambda}(\phi-\m-t)^2
\theta(\phi-\m-t).
\ee
By minimizing $V(\phi)$ we find the $Z_2$ order parameter $\langle
\phi\rangle$, and then use (\ref{psi2}) to obtain the superfluid order
parameter $\langle |\psi|^2\rangle$.

\begin{figure}
\begin{center}
\includegraphics[width=8cm]{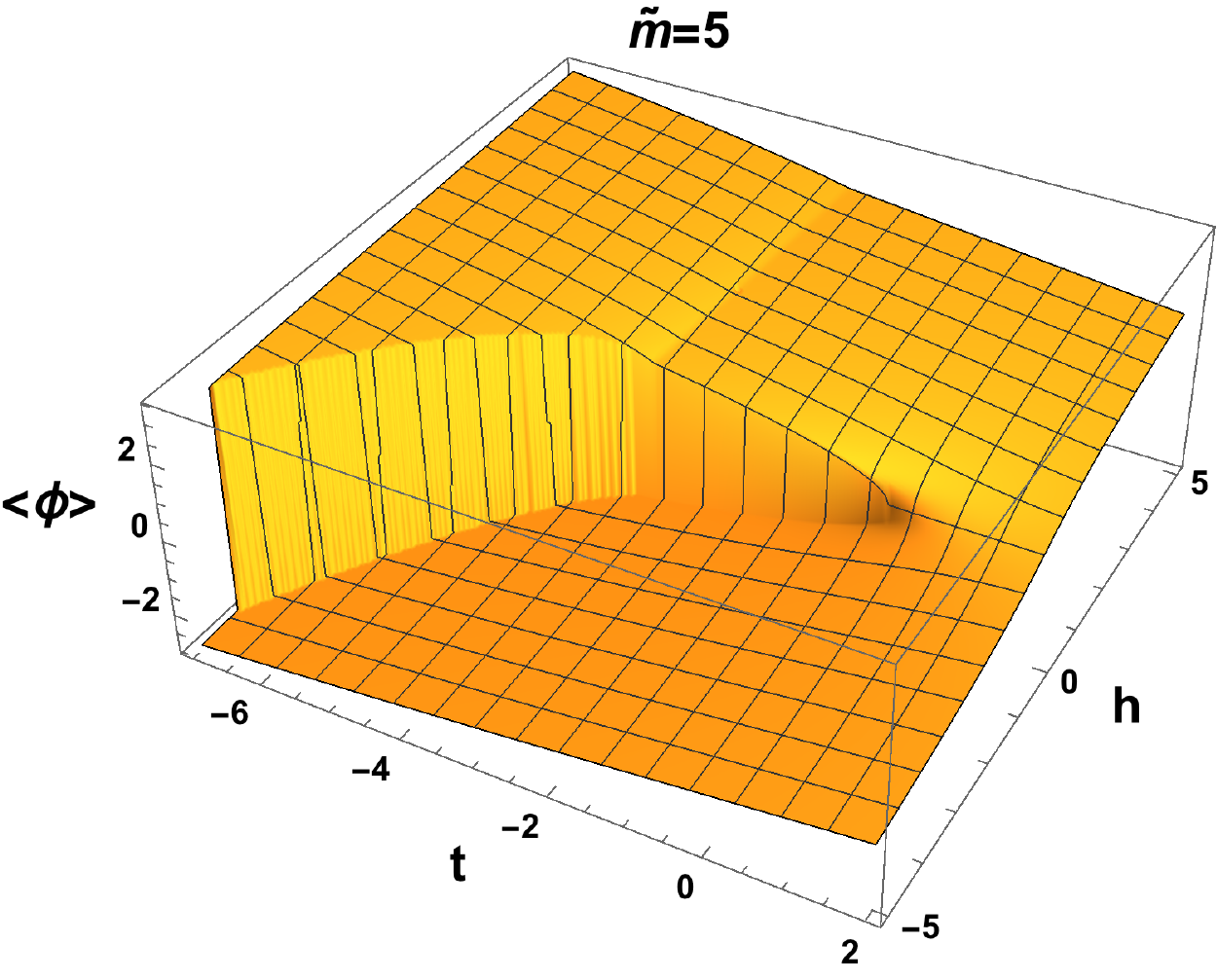}\quad\includegraphics[width=8cm]{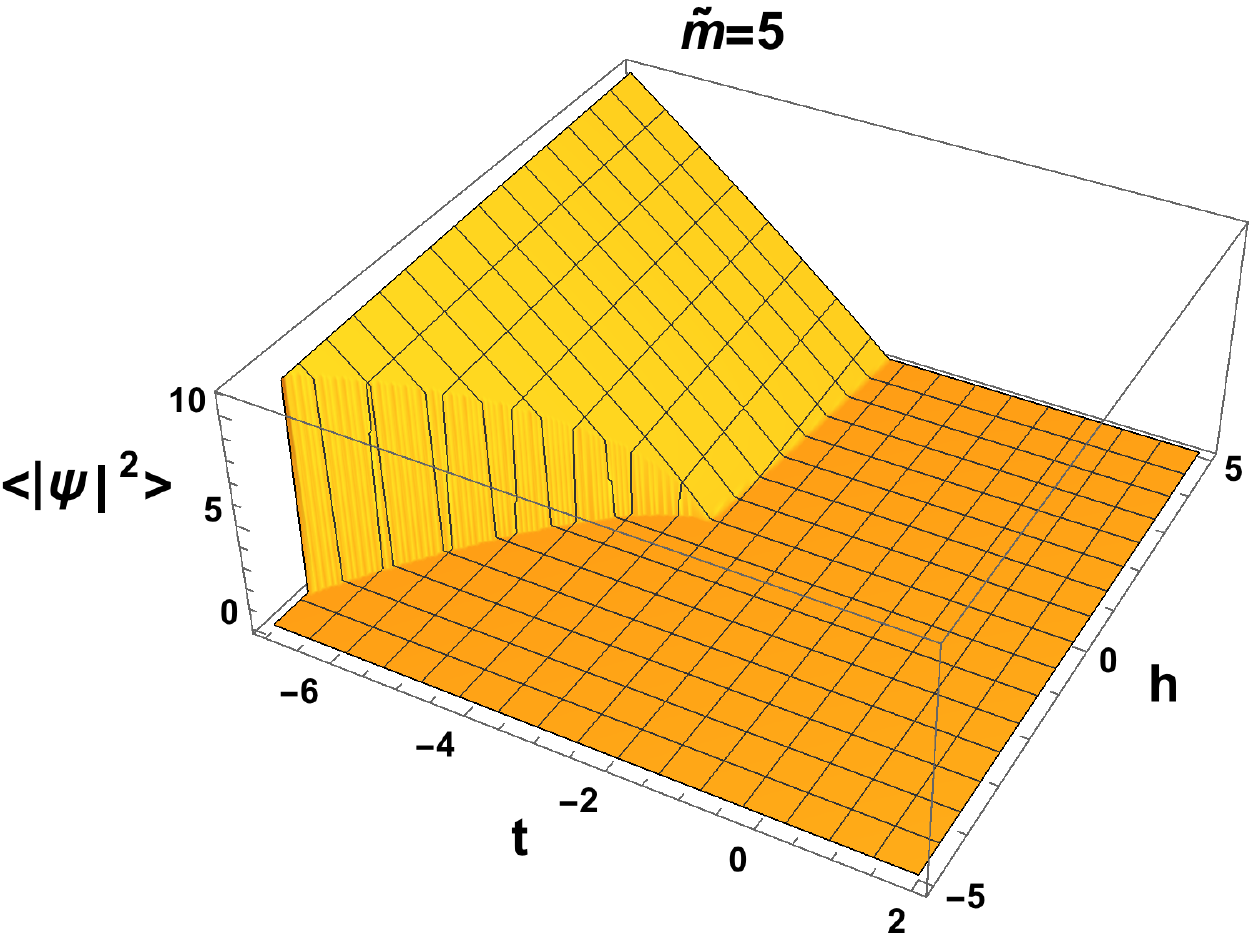}
\includegraphics[width=8cm]{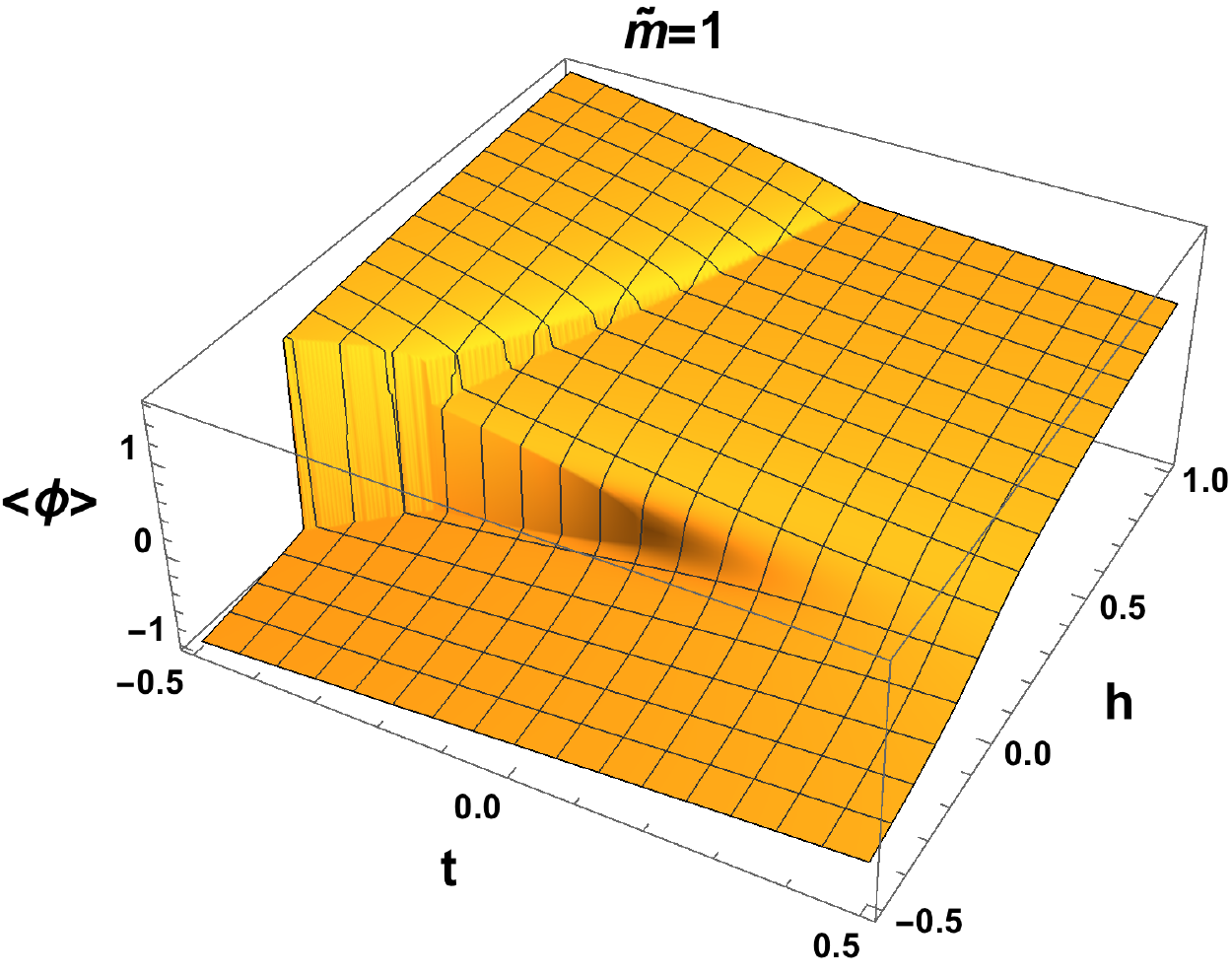}\quad\includegraphics[width=8cm]{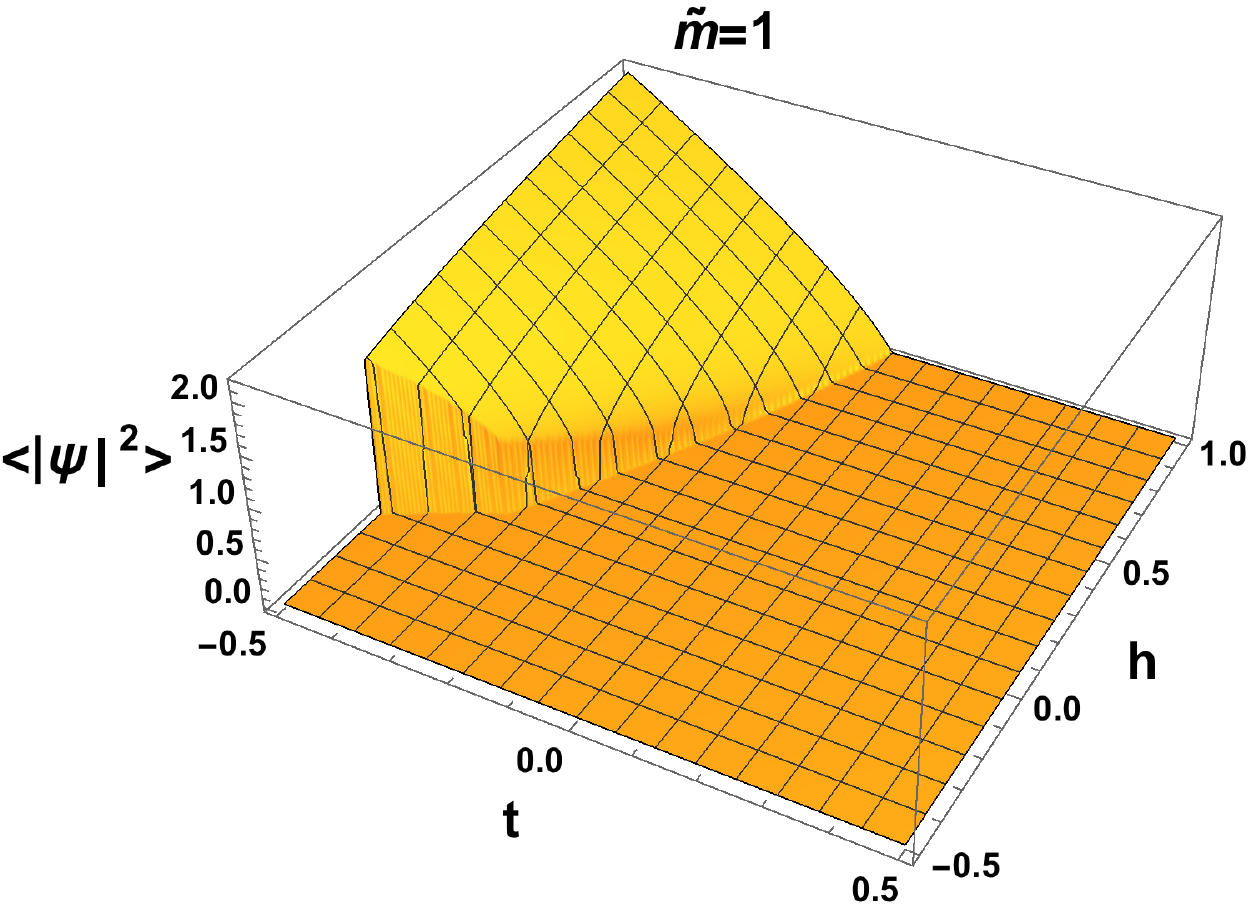}
\includegraphics[width=8cm]{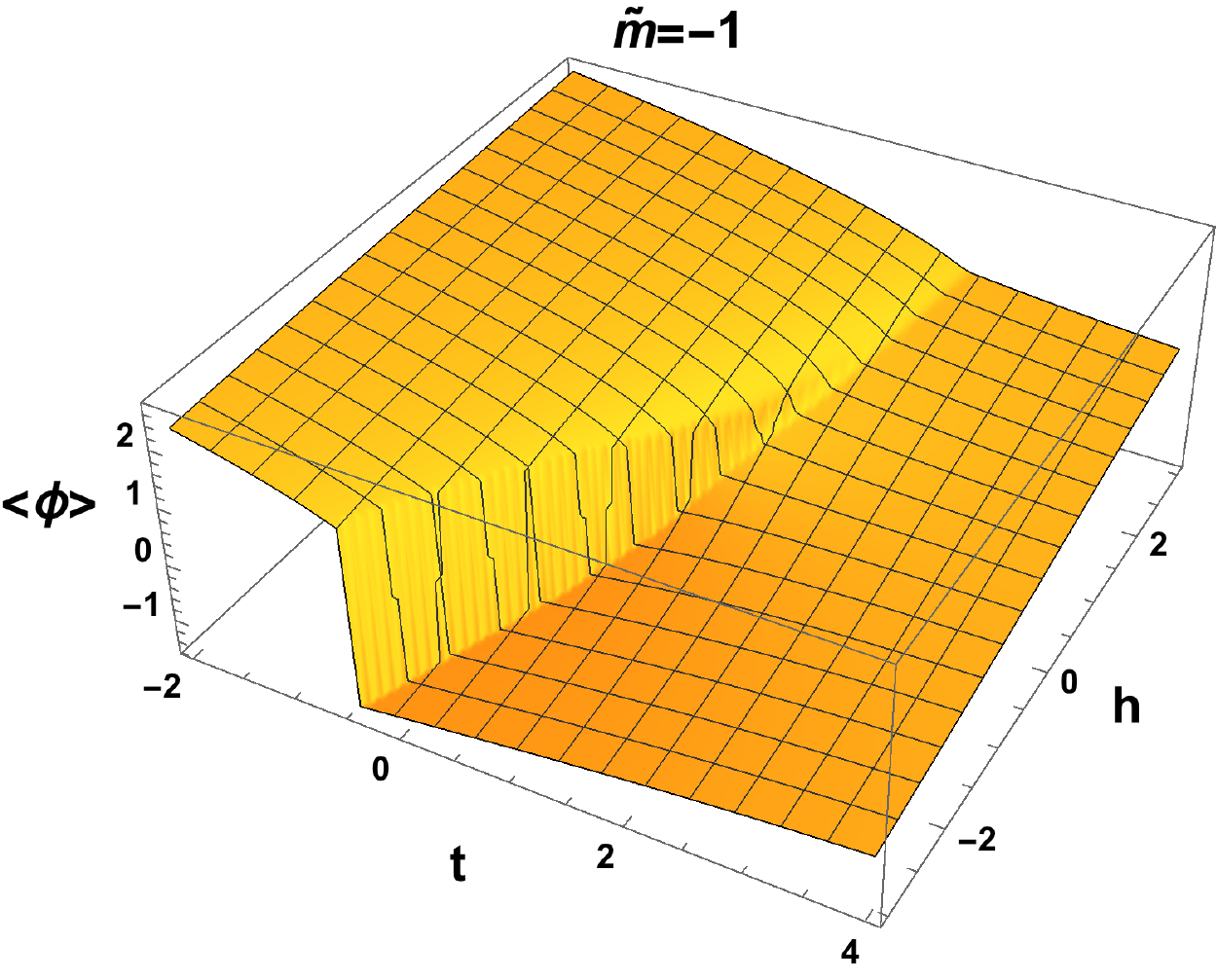}\quad\includegraphics[width=8cm]{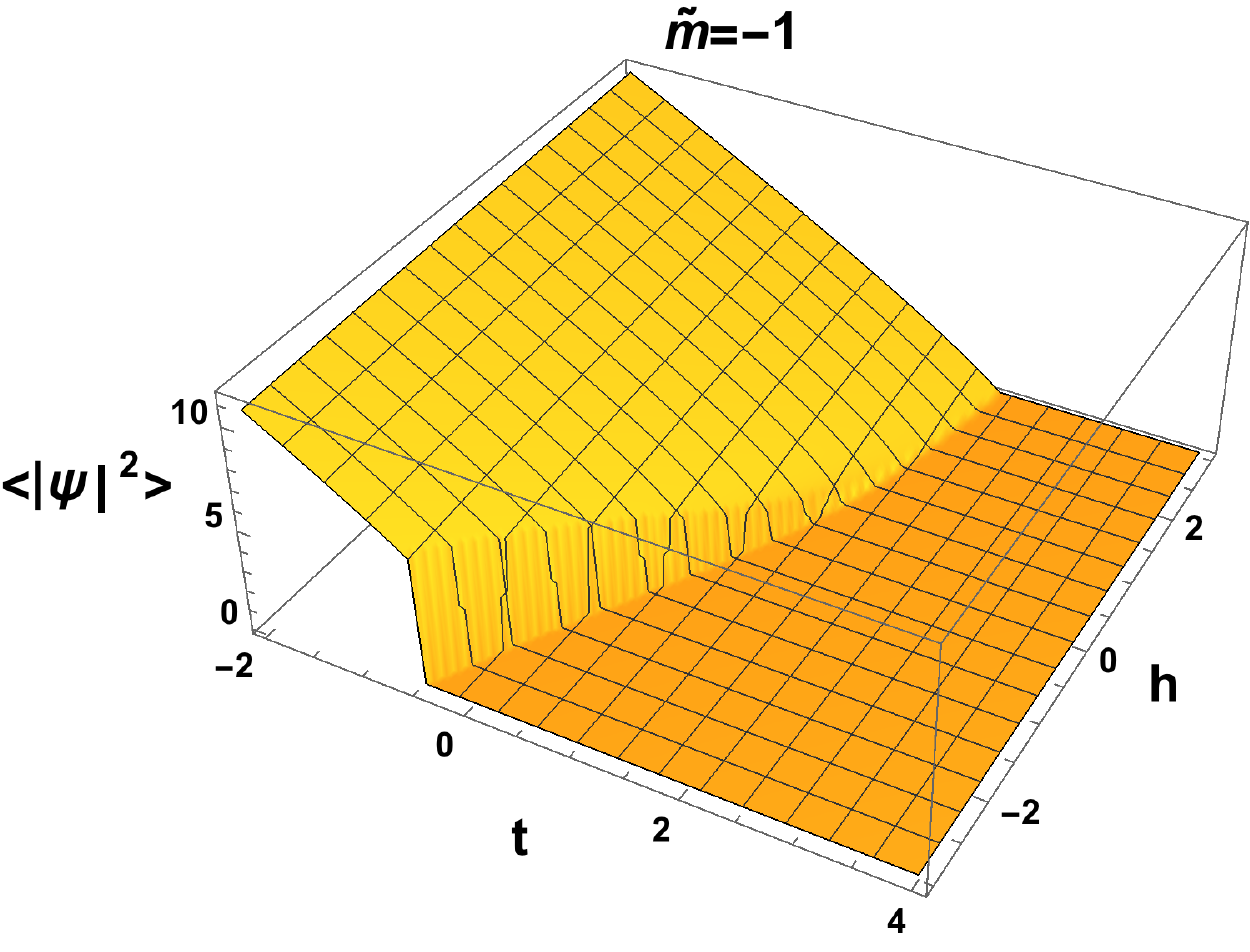}
\caption{The order parameters $\langle\phi\rangle$ (left) and $\langle |\psi|^2\rangle$ (right) in $(t,h)$ space, for $u=1$, $\lambda=1/2$ and $\m=5$, $1$ and $-1$}
\label{fig:numerics}
\end{center}
\end{figure}

The 3D plots on Fig.~\ref{fig:numerics}  show the values of
each order parameter in the $(t,h)$ space, $\langle \phi\rangle
(t,h)$ and $\langle |\psi|^2\rangle (t,h)$.  In the plots one can see
discontinuous jumps represent first-order phase transition lines.
The second-order phase transition lines for superfluid order can also
be seen.  We can play with changing the parameter $\m$ to see how the
phase transition lines change.

The numerical result can be summarized as follows.  For sufficiently
large $\m$, that is $\m>\m_1$, the phase diagram is similar to that of the
physical $^4$He.  As $\m$ drops below $\m_1$, a modification appears
on the line of the superfluid phase transition: the lower part of the
line (which touches the liquid-gas phase transition line) becomes
first order, the upper part remains second order. The two parts are
separated by a tricritical point.  As $\m$ drops further, when
$\m<\m_2$ liquid-gas critical point disappears under the first-order
superfluid-to-normal phase transition line (Fig.~\ref{fig:phasediagram2}).

\subsection{The appearance of the superfluid tricritical point}

One can understand qualitatively why the superfluid phase transition
becomes first-order when this transition happens close to the
liquid-gas phase transition in the following manner.  Suppose we
expand the potential around the minimum at $\phi_0$, i.e.,
$\phi=\phi_0+\delta\phi$.  Then the potential contains the following terms:
\begin{equation}
  V(\phi,\psi) = \cdots + \frac1{2\chi} (\delta\phi)^2 - \delta\phi |\psi|^2
   + \frac\lambda2 |\psi|^4\,,
\end{equation}
where $\chi$ is the susceptibility of the system with respect to the
order parameter $\phi$.  Let us ``integrate out'' the fluctuations of
the liquid-gas order parameter $\phi$ to obtain an effective potential
for $\psi$.  One then finds
\begin{equation}
  V_{\rm eff}(\psi) = \cdots + \left( \frac\lambda2- \frac\chi{2}\right) |\psi|^4 .
\end{equation}
As one approaches the critical point, the susceptibility $\chi\to\infty$
and  $|\psi|^4$ coefficient turns negative.  This
means that the superfluid phase transition will have to become a
first-order phase transition.  In other words, the coupling between
the superfluid order parameter and the liquid-gas order parameter
drives the superfluid phase transition first-order when the
latter occurs near the liquid-gas critical point.

One can use the argument above to locate the point $\m=\m_1$ where
the superfluid tricritical point starts to appear.  At this $\m$ the
tricritical point is located on the line of the liquid gas phase
transition, on the liquid side: $t<0$, $h=+0$.  The average value of
the liquid-gas order parameter at this point is
\begin{equation}
  \langle \phi \rangle = \phi_0 = \sqrt{-\frac tu} \,.
\end{equation}
Expanding $\phi=\phi_0+\delta\phi$, the potential becomes
\begin{equation}
  V = (-t) \delta\phi^2 + u\phi_0\delta\phi^3
     + (t+\m-\langle\phi\rangle)|\psi|^2
      + \frac\lambda2 |\psi|^4 - \delta\phi|\psi|^2 .
\end{equation}
Integrating out $\delta\phi$ means that we minimize the potential with
respect to $\delta\phi$ at fixed $\psi$.  We have to solve
\begin{equation}
  \frac{\d V}{\d \delta\phi} = 2(-t) \delta\phi + 3u\phi_0\delta\phi^2 -
  |\psi|^2 = 0 .
\end{equation}
This equation can be solved for small $|\psi|$:
\begin{equation}
  \delta\phi =  \frac{|\psi|^2}{2(-t)} - \frac{3u\phi_0}{8(-t)^3}|\psi|^4 +
    O (|\psi|^6) . 
\end{equation}
Putting this back into the potential, we find the effective potential
for $\psi$:
\begin{equation}
  V_{\rm eff}(\psi) = (\m+t-\phi_0)|\psi|^2
  + \left( \frac\lambda2 - \frac{1}{4(-t)}\right) |\psi|^4
  + \frac{u\phi_0}{8(-t)^3} |\psi|^6 + O(|\psi|^8) .
\end{equation}
At the tricritical point the coefficients of the $|\psi|^2$ and $|\psi|^4$
terms vanish,
which means
\begin{equation}
  t = -\frac1{2\lambda}\,,\qquad
  \m_1 = \frac1{\sqrt{2\lambda u}} + \frac1{2\lambda} \,.
\end{equation}
Note that at this point the coefficient of the $|\phi|^6$ term is
positive, so this {\em is} the tricritical point.  For example, for
$\lambda=\frac12$ and $u=1$, $\m_1=2$, consistent with the numerical
results in Fig.~\ref{fig:numerics}.

\subsection{Disappearance of the liquid-gas critical point}

Now let us find the value of $\m_2$ at which the liquid-gas critical
point is located on the line of the (first-order) superfluid phase
transition.  At the liquid-gas critical point, we have $t=h=0$, since it is always located in the normal phase side where coupling to superfluid plays no role.  The normal
phase $\phi=\psi=0$ must have the same free energy as the superfluid phase.
As discussed above, the free energy in the superfluid phase is the minimum
of the effective potential
\begin{equation}
  V_{\rm eff}(\phi) = \frac u4 \phi^4 - \frac1{2\lambda}(\phi-\m)^2,
\end{equation}
and the system of equations we need to solve is
\begin{equation}
  V_{\rm eff}(\phi) = \frac u4 \phi^4 - \frac1{2\lambda}(\phi-\m_2)^2 = 0,
  \qquad
  V_{\rm eff}'(\phi) = u\phi^3 - \frac1\lambda(\phi-\m_2) = 0.
\end{equation}
The solution is
\begin{equation}
  \phi=\frac1{\sqrt{2u\lambda}} \,, \qquad
  \m_2 = \frac1{2\sqrt{2u\lambda}} \,.
\end{equation}
For example, when $u=1$, $\lambda=1/2$, $\m_2=\frac12$.  Note that
$\m_1>\m_2$, so as one lowers the atomic mass the superfluid tricritical point
appears before the liquid-gas critical point disappears.

Of course, the mean field model cannot be more than just a guide for
us to guess the correct result, which needs to be obtained by a more
rigorous method.  However, as we will see in the next Section, the
phase diagram depicted in Fig.~\ref{fig:phasediagram2} is realized in
a regime where reliable calculations can be performed.  As a direct
transformation from Fig.~\ref{fig:physical_pd} to
Fig.~\ref{fig:phasediagram2} is impossible, the overall picture
suggested by the mean-field analysis appears to be the simplest one
possible.

\section{Final disappearance of the superfluid tricritical point: the unbinding phase transition}

Further evolution of the phase diagram can be determined, approaching
the problem from another end: from the regime of very small nuclear mass.  Here one
expects that  very strong zero-point fluctuations make the liquid
unbound: the density of the zero-temperature superfluid goes to zero
in the limit of zero pressure.  In this case the superfluid-normal
phase boundary should be completely second order, as in a dilute Bose gas
with repulsive interaction.
Thus, the tricritical point in Fig.~\ref{fig:phasediagram2} moves to lower and lower
temperature and pressure as one decreases the nuclear mass, and
completely disappears at some value of the latter.

The transition between the phase diagram topologies sketched on
Fig.~\ref{fig:phasediagram2} and Fig.~\ref{fig:phasediagram3} occurs
when the scattering length characterizing the low-energy scattering
between two atoms crosses zero~\cite{Sawada:1966,Zwerger:2019}.  For
the Lennard-Jones potential, as the de Boer parameter $\Lambda$
increases, the scattering length $a(\Lambda)$ goes through a series of
poles and zeros (see, e.g., Ref.~\cite{Gomez:2012}).  The last pole of
$a(\Lambda)$ occurs at $\Lambda\approx0.423$; the physical mass of the
$^4$He isotope is very close to this value.  The largest zero of
$a(\Lambda)$ is $\Lambda\approx0.679$, and this corresponds to the
nuclear mass of about $1.58$~u.
Using the more realistic Aziz potentials~\cite{Aziz1979,Aziz1987},
we obtain a slightly lower numerical value of 1.55~u
for the nuclear mass at which $a$ vanishes.

As $m$ approaches $m_3$ from above, the tricritical point moves to
zero temperature and zero pressure.
The density of the superfluid
and the normal gas phases on the two sides of the first-order
section of the superfluid phase transition is small,
therefore one can use effective field theory to describe this phase
transition.

\subsection{Dilute liquid at zero temperature}

At zero temperature, the problem of the dilute droplets of bosons with
small two-body coupling and finite three-body coupling has been
considered in Refs.~\cite{Bulgac2002,Zwerger:2019}.  (Dilute quantum
droplets have recently been considered in the context of trapped
bosons~\cite{Petrov2014,Petrov2015,Chomaz2016,FerrierBarbut2016,Cabrera2017},
but unlike our system,
these droplets are stabilized by effects beyond mean field.)  For completeness, we
rederive the relevant formulas here.

At zero temperature the free energy density, as a function of particle
number density $n$, is given by
\begin{equation}\label{zeroT-Omega}
  \frac\Omega V = \frac g2 n^2 + \frac G 6 n^3 - \mu n ,
\end{equation}
where
\begin{equation}
  g = \frac{4\pi\hbar^2 a}m \,, \qquad G = \frac{\hbar^2D}{m} \,,
\end{equation}
$a$ is the scattering length and $D$ is the three-body scattering
hypervolume~\cite{Tan:2008,Zhu:2017etr}.  The scattering length $a$
approaches zero when the de Boer parameter $\Lambda$ approaches the critical
value $\Lambda_{\rm c}\approx0.679$ as~\cite{Mestrom2020}
\begin{equation}
  a \approx
  3.82812\, \ell_{\rm vdW} (\Lambda-\Lambda_{\rm c})
  \approx
  2.23 \left( \frac\Lambda{\Lambda_{\rm c}} -1 \right)\sigma .
\end{equation}
We will work in the regime $a<0$, $|a|\ll\sigma$.
The three-body scattering hypervolume $D$ has been computed for the
Lennard-Jones potential at $\Lambda=\Lambda_{\rm
  c}$~\cite{Mestrom2020}
\begin{equation}\label{eq:D}
  D =
  (86\pm2)\, \ell_{\rm vdW}^4
  \approx 47 \sigma^4 .
\end{equation}
For $g<0$ a
first-order phase transition occurs at the value of $\mu$ where local
minimum of the free energy~(\ref{zeroT-Omega}) is equal to the free
energy of the vacuum, i.e., zero:
\begin{align}
  - |g| n + \frac G2 n^2 - \mu &= 0, \\
  - \frac{|g|}2 n^2 + \frac G6 n^3 - \mu n &= 0.
\end{align}
These equations have the solution
\begin{align}
  n_0 &= \frac32 \frac{(-g)}G = 6\pi \frac{(-a)}D \,, \label{n0} \\
  \mu_0 &= - \frac38 \frac{g^2}G = - \frac{6\pi^2\hbar^2a^2}{mD}\label{n} \,.
\end{align} 
These formulas coincide with those obtained
Refs.~\cite{Bulgac2002,Zwerger:2019}.

\subsection{Field-theory calculation at finite temperature}

We now investigate the phase diagram of the Bose gas for small and negative
$g$, concentrating on the first-order superfluid phase transition.
We start with the Euclidean Lagrangian for a non-relativistic boson
with chemical potential $\mu$
\begin{equation}
   {\cal L}_E=\psi^\dagger \partial_\tau \psi
   -\frac{\hbar^2 }{2m}|\bm{\nabla}\psi|^2
   -\mu\psi^\dagger \psi +{g\over 2}( \psi^\dagger \psi)^2 
  + \frac{g'}2 |\bm\nabla(\psi^\+\psi)|^2
   + \frac G6 (\psi^\dagger\psi)^3
 .
   \label{Lagrangian}
 \end{equation}
Here in addition to the two- and three-point interactions with
coupling constants $g$ and $G$, we have introduced an interaction
involving extra two spatial derivatives with strength~$g'$.
By matching the scattering amplitude obtained from the Lagrangian
with the effective range expansion, one finds that $g'$ is
proportional to the effective range $r_0$ of the potential,
\begin{equation}
  g' = \frac{\pi\hbar^2}m a^2 r_0.
\end{equation}
Note that when $a\to0$, generically $r_0$ diverges as $a^{-2}$ so that $g'$ remains finite~\footnote{{This can be seen from the formula~\cite{Newton}
$$
    r_0 a^2 = 2\int_0^\infty\!dr\, [(r-a)^2 - u^2(r)],
$$
where $u(r)$ is the solution to the radial Schr\"odinger equation
$-u''(r) + \hbar^{-2}mV(r) u(r)=0$ which vanishes at $r=0$ and tends to $r-a+o(r)$ as  
$r\to\infty$.  This is also how we have obtained the numerical
estimate~(\ref{a2r0-num}) for $r_0a^2$}}.
From dimensional analysis we expect $a^2 r_0\sim\sigma^3$.
Numerically, we find for the Lennard-Jones potential
at $\Lambda=\Lambda_{\rm c}$,
\begin{equation}\label{a2r0-num}
  \lim_{\Lambda\to\Lambda_{\rm c}} a^2 r_0 \approx 5.13 \, \ell_{\rm vdW}^3 \approx
  3.23 \, \sigma^3.
\end{equation}
Normally, at small density the interaction term proportional to $g$ dominates, but
since the coefficient of this term, $g$, is tuned to zero, one needs
to take into account interactions which involve more particles ($G$)
or more derivatives ($g'$).   As we shall discuss in more detail
below, in the regime of our interest, the addition of a pair of operators $\psi^\+$ and $\psi$
brings a suppression factor of $|a|/\sigma$, and a pair of spatial derivatives
a factor of $(|a|/\sigma)^{2/3}$.

We expand the field in Fourier modes of Matsubara frequency and momentum,
\be
\psi(\tau,\bm x)=
\sum_{n\in Z}e^{-{2\pi i n\tau\over \beta}}\psi_{n}(\bm x)
=\sum_{n\in Z}e^{-{2\pi i n\tau\over \beta}}\int_{\bm k}e^{i\bm k\cdot\bm x} \psi_{n,\bm k} ,
\ee
where $\beta\equiv 1/T$ and 
\be
\int_{\bm k}\equiv \int{d^3\bm k\over (2\pi)^3}\,.
\ee
The Euclidean action then has the form
\bear
S_E/\beta&=&\int_{\bm k}\sum_n \psi^*_{n,\bm k}\left(-{2\pi i n\over\beta}+{\hbar^2\bm k^2\over 2m}-\mu\right)\psi_{n,\bm k}\nonumber\\
&+&{g\over 2}\int_{\bm x}\sum_{n_1,n_2,n_3,n_4} \psi^*_{n_1}\psi^*_{n_2}\psi_{n_3}\psi_{n_4}\delta_{n_1+n_2-n_3-n_4}e^{{2\pi i \varepsilon\over\beta}(n_1+n_2)}\nonumber\\
&+& {g'\over 2}\int_{\bm x}\sum_{n_1,n_2,n_3,n_4} \bm\nabla_i(\psi^*_{n_1}\psi_{n_3})\bm\nabla_i(\psi^*_{n_2}\psi_{n_4})\delta_{n_1+n_2-n_3-n_4}e^{{2\pi i \varepsilon\over\beta}(n_1+n_2)}
\nonumber\\
&+&{G\over 6}\int_{\bm x}\sum_{n_1,n_2,n_3,n_4,n_5,n_6}\psi^*_{n_1}\psi^*_{n_2}\psi^*_{n_3}\psi_{n_4}\psi_{n_5}\psi_{n_6}\delta_{n_1+n_2+n_3-n_4-n_5-n_6}e^{{2\pi i \varepsilon\over\beta}(n_1+n_2+n_3)} ,
\eear
where the last factor with $\varepsilon=0^+$ is from the time ordering of $\psi^\dagger(\tau+\varepsilon)\psi(\tau)$ in the interaction terms that ensures the normal ordering of $\psi^\dagger \psi$ in the operator form.

One can now derive the Feynman rules for the theory.  The free
propagator is given by
\be
\langle \psi^*_{n',\bm k'}\psi_{n,\bm k}\rangle={1\over -{2\pi i n}+\beta\left({\hbar^2\bm k^2\over 2m}-\mu\right)}(2\pi)^3\delta(\bm k'-\bm k)\delta_{n'-n} .
\ee

The interaction terms in the Lagrangian in Eq.~(\ref{Lagrangian})
generate three types of vertices: a four-point vertex $-2g\beta$, a
four-point vertex equal to $g'\beta$ multiplied by
\begin{equation}
2 (\bm p_1 \cdot \bm p_2 + \bm k_1\cdot \bm k_2)
  -(\bm p_1 + \bm p_2) \cdot (\bm k_1 + \bm k_2)=-(\bm p_1-\bm k_1)^2-(\bm p_1-\bm k_2)^2\,,
\end{equation}
where the incoming momenta are denoted as 
$\bm p_1$, $\bm p_2$ and outgoing momenta as $\bm k_1$, $\bm k_2$,
and a six-point vertex $-6G\beta$.

\begin{figure}
\begin{center}
\includegraphics[width=9cm]{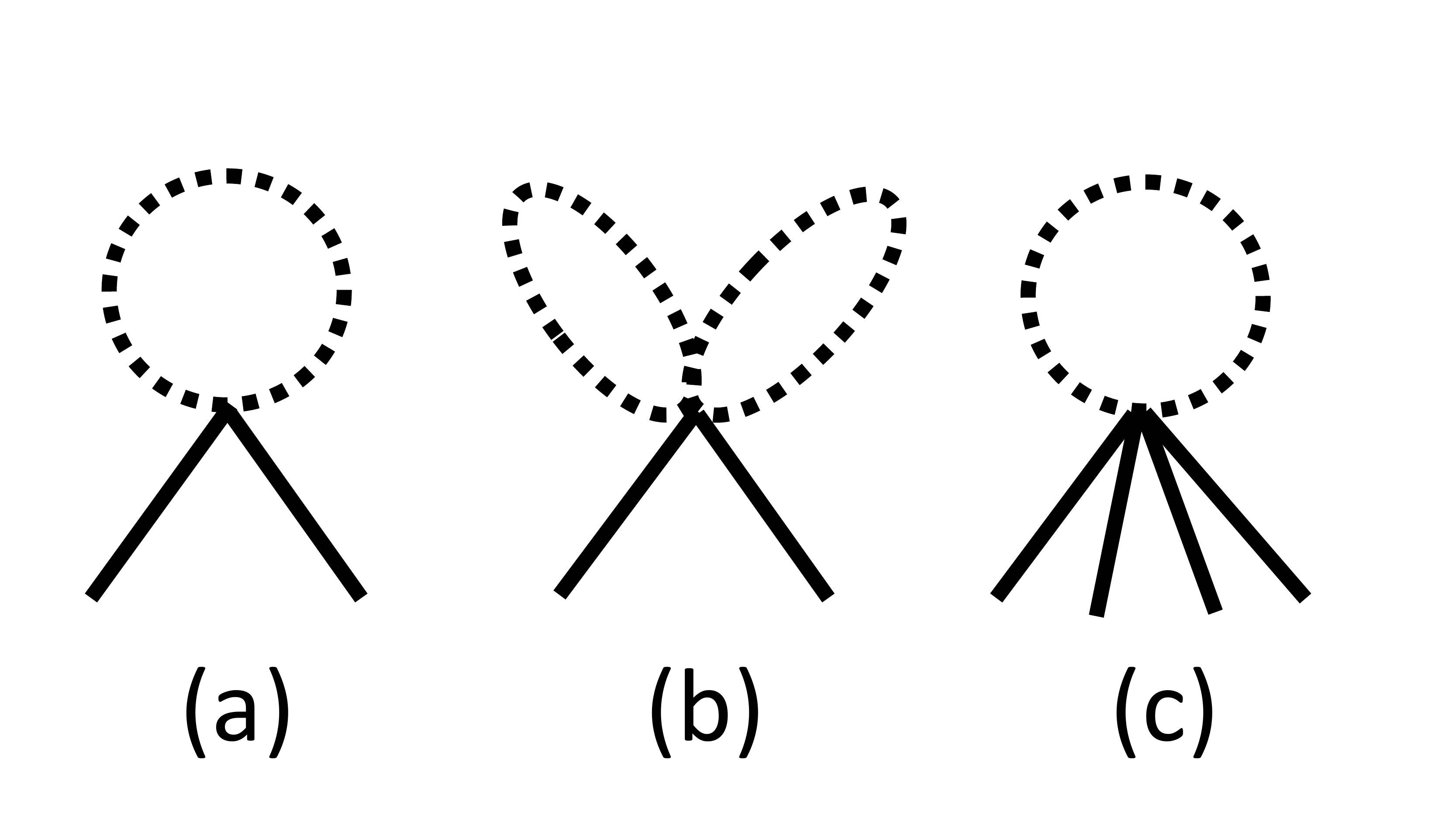}
\caption{The ``tadpole" diagrams contributing to effective parameters
  in  the 3D effective theory. Diagrams (a) and (b) contribute to
  $\mu_{\rm eff}$, diagram (a) with $g'$ vertex to
  $m_{\rm eff}$, and diagram (c)   to $g_{\rm eff}$ (see Eq.~(\ref{mug-eff})). Dashed lines are $n\neq 0$ modes that are integrated out.}
\label{fig:tadpole}
\end{center}
\end{figure}

We integrate out $n\neq 0$ modes to obtain a three-dimensional effective
field theory for the zero Matsubara frequency ($n=0$) modes. 
As shown below and in the Appendices, the small parameter controlling
the loop expansion is the ratio of the scattering length $a$
to the range of the Lennard-Jones potential~$\sigma$:
\begin{equation}
  \label{eq:as}
  \frac{|a|}{\sigma}\ll1\,.
\end{equation}
There are two other relevant length parameters: the typical
interparticle spacing $n^{-1/3}$ and the thermal length $\lambda_T$,
given by
\begin{equation}\label{eq:lambda}
 \lambda_T = \sqrt{2\pi\hbar^2\over mT}\,.
\end{equation}
However, in the regime we want to consider, i.e.,
densities of order $n_0$ and temperatures of order the tricritical temperature,
both $n_0$ and $\lambda_T$ are given in terms of $a$ and the small parameter in
Eq.~(\ref{eq:as}).  In particular, the density in the regime where we
are working is of order $n_0$ given in Eq.~(\ref{n0}), and since, on
dimensional grounds, the hypervolume $D$ is of order~$\sigma^4$~(see Eq.~(\ref{eq:D})), we see
that $|a|\ll\sigma$ corresponds to the dilute limit:
\begin{equation}
  \label{eq:na}
  n |a|^3 \sim \frac{a^4}{D} \sim \left(\frac{a}{\sigma}\right)^4\ll1. 
\end{equation}
The thermal length $\lambda_T$ at the temperatures we are interested in,
i.e., near the Bose-Einstein condensation transition, is of order the
particle separation $n^{-1/3}$, and is thus also set in terms of the
ratio in Eq.~(\ref{eq:as}).

We find that in the regime $|a|\ll\sigma$ the tadpole diagrams
dominate over other loop diagrams, as far as the determination of the tricritical
point at $T_{\rm tri}$ and the first-order phase transition line for
$T<T_{\rm tri}$ is concerned.
(We use the term ``tadpole" for a diagram with an internal line which
starts and ends at the same point, as in Fig.~\ref{fig:tadpole}). 
Physically, the tadpoles correspond to thermal occupation of bosonic excitations above
the condensate (the
$n=0$ mode).  Integrating out tadpoles with $n\neq0$ modes corresponds to
incorporating the effects of the normal component
on the dynamics of the superfluid condensate.

In Appendices~\ref{sec:supr-therm-non}
and~\ref{sec:suppression-loops-3d} we show that non-tadpole diagrams
with nonzero Matsubara frequency modes and the loop diagrams within
the 3D effective field theory can be neglected in the regime we are
considering, i.e., near the tricritical point for a system with
$|a|\ll\sigma$.

There are two tadpole diagrams contributing to the corrections to the 3D effective parameter $\mu_{\rm eff}$, each coming from four-point
and six-point interaction vertices.
The contribution from the tadpoles with a four-point vertex is 
\be\label{sum-subtracted}
\Delta \mu=-\!\int_{\bm k}(2g + g' \bm k^2)\sum_{n\neq 0}{e^{2\pi i n\varepsilon\over\beta}\over -{2\pi i n}+\beta\left({\hbar^2\bm k^2\over 2m}-\mu\right)}
\,.
\ee
The numerator in Eq.~(\ref{sum-subtracted}) comes from the normal ordering of operators in the interaction vertices. If $n$ summation included $n=0$, the result would be simply the Bose-Einstein distribution function,
\be
\sum_{n\in Z}{e^{2\pi i n\varepsilon\over\beta}\over -{2\pi i n}+\beta\left({\hbar^2\bm k^2\over 2m}-\mu\right)}= {1\over \exp\left[{\beta\left({\hbar^2\bm k^2\over 2m}-\mu\right)}\right]-1}\,,
\ee
so the sum in Eq.~(\ref{sum-subtracted}) is simply this with $n=0$ term subtracted,
\be
\Delta\mu=- \int_{\bm k}(2g + g' \bm k^2) \left({1\over \exp\left[{\beta\left({\hbar^2\bm k^2\over 2m}-\mu\right)}\right]-1}-{1\over \beta\left({\hbar^2\bm k^2\over 2m}-\mu\right)}\right).
\ee
The first integral over $\bm k$ is finite,
but the second term is UV divergent.  This divergence should cancel with
the UV divergences in the loop diagrams of the 3D effective theory if
the same regularization scheme is used in both calculations.
It is convenient to use dimensional
regularization, in which tadpoles do not have UV divergences.
We have
\be
\Delta\mu= - {2g\over \lambda_T^3} L(\beta\mu)
   - \frac{6\pi g'}{\lambda_T^5} M(\beta\mu),
\ee
where
\begin{align}
  L(x) &= \Li_{3/2}\left( e^{x}\right)+\sqrt{-4\pi x}\, ,\\
  M(x) &= \Li_{5/2}\left( e^x \right) - \frac43 \sqrt{-\pi x^3} \,,
\end{align}
and the thermal wavelength $\lambda_T$ is defined in Eq.~(\ref{eq:lambda}).

The tadpole contribution to $\mu_{\rm eff}$ from the six-point vertex is similarly computed to be
\be
\Delta \mu=-{3G\over\lambda_T^6}
L^2(\beta\mu),
\ee
and the tadpole contribution to $g_{\rm eff}$ from the $G$-vertex is
\be
\Delta g={3G\over \lambda_T^3}
L(\beta\mu)\,.
\ee
There is also a correction to the mass $m$:
\begin{equation}
  \Delta \left(\frac1 m\right) =
  \frac{2g'}{\hbar^2\lambda_T^3} L(\beta\mu)
\end{equation}

Thus, the 3D effective theory is described by the Lagrangian
\begin{equation}
{\cal L}_{\rm 3D}/\beta= {\hbar^2\over 2m_{\rm eff}}|\bm{\nabla}\psi_0|^2-\mu_{\rm eff}\psi^\dagger_0 \psi_0 +{g_{\rm eff}\over 2}( \psi^\dagger_0 \psi_0)^2 +{G\over 6}(\psi^\dagger_0\psi_0)^3, \label{3Deff}
\end{equation}
with the parameters
\begin{subequations}\label{mug-eff}
\begin{align}
\mu_{\rm eff} &= \mu
-{2g\over \lambda_T^3} L(\beta\mu)
 - \frac{6\pi g'}{\lambda_T^5} M(\beta\mu)
-{3G\over\lambda_T^6} L^2(\beta\mu) , \label{mueff}
\\\label{eq:geff}
g_{\rm eff} &= g+{3G\over \lambda_T^3} L(\beta\mu), \\\label{eq:meff}
\frac1{m_{\rm eff}} &= \frac1m + \frac{2g'}{\hbar^2\lambda_T^3} L(\beta\mu)\,.
\end{align}
\end{subequations}
As we show in Appendix \ref{sec:suppression-loops-3d}, the
contributions of
higher-derivative and higher-order terms are suppressed in the
effective theory given by Eq.~(\ref{3Deff}).
The tadpole contributions in Eq.~(\ref{mueff}) are described by
diagrams in Fig.~\ref{fig:tadpole}(a) with $g$ and $g'$ vertex, as
well as the diagram in Fig.~\ref{fig:tadpole}(b) with $G$ vertex
correspondingly. The tadpole contribution in Eq.~(\ref{eq:geff}) is
from the diagram in Fig.~\ref{fig:tadpole}(b), and the tadpole
contribution in Eq.~(\ref{eq:meff}) is from the diagram in
Fig.~\ref{fig:tadpole}(a) with $g'$ vertex.

The first order phase transition line is determined by solving the equation for $\mu(T)$,
\be
\mu_{\rm eff}= - {3g_{\rm eff}^2\over 8G} \,,
\ee
and the tricritical point is at $g_{\rm eff}=\mu_{\rm eff}=0$.

We will see that the most interesting regime is $|\beta\mu|\ll1$ ($\beta\mu<0$).
In this regime, the functions $L(x)$ and $M(x)$ tend to constant values
\begin{subequations}\label{L-series}
\begin{align}
L(\beta\mu) &= \zeta\left(\frac32 \right) + \zeta\left(\frac12\right)\beta\mu+\cdots \,,\\
M(\beta\mu) &= \zeta\left(\frac52 \right) + \zeta\left(\frac32\right)\beta\mu+\cdots \,.
\end{align}
\end{subequations}
Let us define
\begin{equation}
  n_{\rm B}(T) \equiv\zeta\left(\frac32\right) \frac1{\lambda_T^3} =
   \zeta\left(\frac32\right) \left( \frac{mT}{2\pi\hbar}\right)^{3/2}\,,
\end{equation}
which is the density of a free Bose gas with temperature $T$ and zero chemical potential.
Now Eqs.~(\ref{mug-eff}) become
\begin{align}
  \mu_{\rm eff} &= \mu - 2g n_{\rm B}(T) - 3G n_{\rm B}^2(T) 
  - \frac{6\pi\zeta(\frac52)}{\zeta(\frac32)} \frac1{\lambda_T^2} g' n_{\rm B}(T)
  \,, \label{mueffnB}\\
  g_{\rm eff} &= g + 3 G n_{\rm B}(T)\,.
\end{align}

\subsection{The tricritical point}

The phase transition is first order when $T<T_{\rm tri}$ and second
order when $T>T_{\rm tri}$, where the tricritical temperature $T_{\rm
  tri}$ is determined by requiring $g_{\rm eff}=0$:
\begin{equation}\label{Ttri}
  T_{\rm tri} = \frac{2\pi\hbar^2}m
  \left[ \frac{(-g)}{3\zeta(\frac32)G}\right]^{2/3} =
  \frac{2\pi\hbar^2}m \left[ \frac{4\pi(-a)}{3\zeta(\frac32)D}\right]^{2/3} .
\end{equation}
The chemical potential at the tricritical point is determined from
$\mu_{\rm eff}=0$.  From Eqs.~(\ref{mueffnB}) and (\ref{Ttri}) it follows that
\begin{equation}\label{mutrigg}
  \mu_{\rm tri} =
  6\pi \zeta\left( \frac52 \right) g'
  \left( \frac{-g}{3\zeta(\frac32)G}\right)^{5/3}
  -\frac13 \frac{g^2}G \,.
\end{equation}

In the limit $g\to0$, the first term on the right-hand side of
Eq.~(\ref{mutrigg}) dominates over the second term, and one can estimate
\begin{equation}
  \mu_{\rm tri} \sim \frac{\hbar^2|a|^{5/3}}{m \sigma^{11/3}} \,.
\end{equation}

One can now check that the condition $\beta\mu\ll1$ is satisfied at
the tricritical point:
\begin{equation}\label{betamutri}
  \beta \mu \sim \frac{|a|}{\sigma} \ll 1.
\end{equation}
The correction to the mass is also small:
\begin{equation}\label{Deltamtri}
  \Delta \left( \frac1m \right) \sim \frac{g'}{\hbar^2\lambda_T(T_{\rm tri})}
  \sim \frac1m \frac{|a|}\sigma \ll \frac1m
\end{equation}
One expects that (\ref{betamutri}) and (\ref{Deltamtri}) are valid not
only at the tricritical point, but also in the whole regime of
temperature and chemical potential $T\sim T_{\rm tri}$,
$\mu\sim\mu_{\rm tri}$.

Since $|\mu|\ll T$, the particle number at the
tricritical point is given by
\begin{equation}
  n_{\rm tri} = n_{\rm B}(T_{\rm tri}) = \frac13 \frac{(-g)}{G}\,.\label{ntri}
\end{equation}
This is equal to $\frac29$ of the density of the liquid phase at zero
temperature $n_0$ [Eq.~(\ref{n0})].

\subsection{The first-order phase transition}
\label{sec:first-order-phase}

For $T<T_{\rm tri}$, the phase transition is first-order.  Assuming
that $T\sim T_{\rm tri}$, then $\mu\sim\mu_{\rm tri}$ and the condition
$\beta\mu\ll1$ is still valid.  The chemical potential at the
first-order phase transition is determined via
\begin{equation}
  \mu_{\rm eff} = -\frac38 \frac{g_{\rm eff}^2}G\,.
\end{equation}

Let us now compute the density along the coexistence curve.  In the
gas phase the density is equal to the density of a Bose gas with
temperature $T$ and zero chemical potential
\begin{equation}\label{eq:ngas}
  n_{\rm gas} = n_{\rm B}(T) = \left( \frac T{T_{\rm tri}}\right)^{3/2}  n_{\rm tri} = \frac29 \left( \frac T{T_{\rm tri}}\right)^{3/2} n_0  .
\end{equation}
The density in the (superfluid) liquid phase is the sum of the
condensate density
and the density of the thermal excitations:
\begin{equation}\label{eq:nliq}
  n_{\rm liq} = n_{\rm cond} + n_B(T),
\end{equation}
where the condensate density is
\begin{equation}
  n_{\rm cond} = \frac32 \frac{(-g_{\rm eff})}G =
  \frac32 \frac{(-g)}G - \frac92 n_{\rm B}(T) =
  \frac92 \left[1 - \left( \frac{T}{T_{\rm tri}}\right)^{3/2}\right]n_{\rm tri}
  .
\end{equation}
Therefore
\begin{equation}
  n_{\rm liq} = \left[ \frac92 - \frac72\left( \frac{T}{T_{\rm tri}}\right)^{3/2}  \right] n_{\rm tri}
  = \left[ 1 - \frac79\left( \frac{T}{T_{\rm tri}}\right)^{3/2}  \right] n_{0}
  . \label{nfirst}
\end{equation}

\begin{figure}
  \centering
  \includegraphics{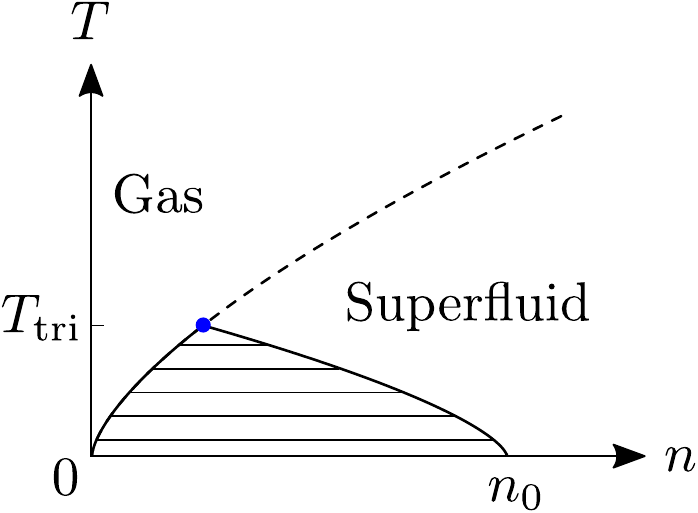}
  \caption{The phase diagram in the temperature vs density plane
    corresponding to Fig.~\ref{fig:phasediagram2}.}
  \label{fig:nT}
\end{figure}

Equations (\ref{eq:ngas}) and (\ref{eq:nliq}) describe the boundaries of
the coexistence region on the $T$ vs $n$ phase diagram as shown in
Fig.~\ref{fig:nT}.

By minimizing the free energy given by Eq.~(\ref{3Deff}) for a configuration interpolating between
the two phases, one finds
the profile of the superfluid order parameter on the interface,
\begin{equation}
  \psi_0(x) =
  \sqrt{\frac{n_{\rm cond}}{1+e^{2x/\xi}}} \,, \qquad
  \xi = \frac{1}{2\pi\sqrt3} \frac{D^{1/2}}{|a|}
  \left[ 1- \left( \frac{T}{T_{\rm tri}}\right)^{3/2} \right]^{-1},
\end{equation}
and the surface tension,
\begin{equation}
  \tau = 3\sqrt3 \pi^2 \frac{\hbar^2 a^2}{m D^{3/2}}
  \left[ 1- \left( \frac{T}{T_{\rm tri}}\right)^{3/2} \right]^{2}.
\end{equation}
The surface tension vanishes as $(T_{\rm tri}-T)^2$ and the thickness of the interface diverges as $(T_{\rm tri}-T)^{-1}$ near $T_{\rm tri}$.

\section{Possible realizations of ultra quantum liquids}

In the real world the mass of the helium nucleus is fixed.  There are
two bosonic isotopes of helium, $^6$He and $^8$He, with half-lives of
0.8~s and 0.12~s, respectively.  The lifetime of these isotopes is
very large compared to the microscopic time scales and thus the question about
the phase diagram of these isotopes make sense.  The nuclear masses of
these isotopes, however, lie on the other side of the mass of the
$^4$He nucleus, compared to the mass region explored in this paper.
The behavior of $^6$He and $^8$He must be more classical than that of~$^4$He.

Substances other than helium have smaller de Boer parameter and hence
are more classical.  Let us define, for a given substance, the
``equivalent helium mass'' to be the mass of a helium isotope (measured
in atomic mass unit) that would have the same de Boer parameter as
that of the chosen substance:
\begin{equation}
  M_{\rm eq} = \frac{m}{m_N}  \frac\epsilon{\epsilon_{\rm He}}
  \left( \frac\sigma{\sigma_{\rm He}} \right)^2 .
\end{equation}
The equivalent helium mass for selected substances are given on the
second row of Table~\ref{table:substances}, where we have used
the Lennard-Jones parameters from Ref.~\cite{BerryRiceRoss}.
\begin{table}
\begin{tabular}{c|c|c|c|c|c|c|c|c|c}
    & $^4$He & $^{20}$Ne & $^{40}$Ar & $^{84}$Kr & H$_2$ & N$_2$ & O$_2$ & CO & CH$_4$\\
    \hline
    $M_{\rm eq}$ & 4 & 80.6 & 832 & 2790 & 9.50 & 545 & 722 & 595 & 517\\
    \hline
    $M_{\rm eq}(\mu)$ & 0.020 & 0.412 & 4.23 & 14.1 & 0.051 & 2.79 & 3.69 & 3.04 & 2.68
\end{tabular}
\caption{Equivalent helium mass for selected substances and their muonic versions.}
\label{table:substances}
\end{table}

One possible (but admittedly experimentally very difficult) way to
achieve a de Boer parameter larger than that of $^4$He is to create
``muonic matter'' by replacing all electrons in a given
substance by by muons~\cite{Tajima:1987,Wheeler:1988}.  The effect of
this replacement is to increase the depth of the potential $\epsilon$
by a factor of the ratio of the muon mass $m_\mu$ to the electron mass $m_e$,
$m_\mu/m_e\approx 207$, and to decrease the range of the
potential $\sigma$ by the same factor.  This has an effect of reducing
the equivalent helium mass
\begin{equation}
  M_{\rm eq}(\mu) = \left(1 + \frac ZA \frac{m_\mu}{m_N} \right)
  \frac{m_e}{m_\mu} M_{\rm eq}(e)
\end{equation}
(we have taken into account a small change of the mass of the atom).
The equivalent helium mass of the muonic substances are given in the
third row of Table~\ref{table:substances}.  One can see that among
the noble gases, muonic argon has essentially the same de Boer
parameter as that of (electronic) $^4$He, and thus will have a phase
diagram very similar to that of $^4$He (with the superfluid transition
temperature of order 5000~K).  At the same time muonic neon would be
deep in the ``gas-like'' phase and muonic krypton and xenon should be
rather classical.  On the other hand, muonic N$_2$, O$_2$, CO and
CH$_4$ have the equivalent helium mass within the interesting range (from
1.55 to 4) and hence can realize the various versions of the phase
diagram treated in this paper.

Another type of exotic matter is formed when one replaces the protons
in H$_2$ by a lighter positively charged particles, for example, muons or pions.  Denoting the mass of the particle that replaces the proton as $m$, the equivalent helium mass of this
substance would be
\begin{equation}
  M_{\rm eq} =2 m \frac{\sigma_{{\rm H}_2}^2 \epsilon_{{\rm H}_2}}
  {\sigma_{{\rm He}}^2 \epsilon_{{\rm He}}}
  \approx 9.50 m \approx \frac1{193} \left( \frac{m}{m_e}\right)\textrm{u} .
\end{equation}
According to this formula, when $m\approx 770 m_e$ the substance would behave like helium, and
when $m< 300 m_e$, its phase diagram is that of a repulsive Bose gas.

Previously, Wheeler has put the estimate of the critical mass
of the proton replacement at which the self-bound liquid
disappears at 253 $m_e$~\cite{Wheeler:1988}.
One can try to determine this value better by using
a more accurate interaction potential between two hydrogen molecules,
for example, the widely used 
Silvera-Goldman potential~\cite{Silvera1978} or the Buck
potential~\cite{Buck1983}.
It turns out that these potentials give results
practically indistinguishable from Wheeler's estimate (250~$m_e$ for the
Silvera-Goldman potential and 256~$m_e$ for the Buck potential).
One can obtain a very good estimate almost
analytically by noticing that the 10-6
potential
\begin{equation}
  V(r) = c_{10-6}\, \epsilon \left[ \left( \frac\sigma r\right)^{10}
    - \left( \frac\sigma r \right)^6 \right]
\end{equation}
(where $c_{10-6}=2^{-1}3^{-3/2}5^{5/2}\approx 5.379$ is chosen so that $\epsilon$ is the potential depth at its minimum) provides a considerably better
approximation to the Silvera-Goldman and Buck potentials,
compared to the 12-6 potential.
The added advantage
of the 10-6 potential is that the scattering length is known
analytically in terms of potential parameters~\cite{Pade:2007};
in particular, the last zero of $a$
occurs at the value of the de Boer parameter
$\Lambda=\hbar/\sigma \sqrt{2m\epsilon}=\sqrt{c_{10-6}}/3\approx0.773$.  Using
the core radius and the potential depth of the Silvera-Goldman potential,
$\sigma=2.974$~\AA{} and $\epsilon/k_B=34.31$~K,
one obtains $m\approx 243\, m_e$, surprisingly close to the result obtained
by numerically integrating the Schr\"odinger equation, given
the crudeness of the approximation.

From the estimate above, it appears that ``muonium matter,'' matter made out of muonium
molecules $(\mu^+e^-)_2$ or Mu$_2$, with $m_\mu/m_e\approx 207$, 
would not form a self-bound liquid and would
have a phase diagram of the type of a repulsive
Bose gas,
while the molecules $(\pi^+e^-)_2$, made out of positive pions and electrons with
$m_{\pi^+}/m_e\approx 273$, is capable of forming a self-bound liquid.
However, given the short lifetimes of the muon and the pion,
perhaps a more realistic way to realize a quantum liquid
of H$_2$-like molecules
is to use excitons in a solid where the hole
and the electron have very different effective masses.  When the ratio of
the masses of the hole and the electron that form the exciton varies
between approximately 250 and 800, the fluid of biexcitons~\cite{Moskalenko}
could realize different versions of the phase diagram discussed in this paper.
This estimate will change if the mass tensors are
anisotropic.

\section{Conclusion}

In this paper we have followed the evolution of the phase diagram of
helium-4 as one dials  the nucleus mass from the physical
value of 4~u to a value where the scattering length vanishes, estimated
to be $1.55$~u.  We have argued that the phase
diagram goes through a sequence of changes before it finally becomes a
quantum ``gas-like'' phase diagram, where there is only a single
second-order phase transition between the superfluid and the normal
phases.

Except for the last change in the phase diagram---the disappearance of
the superfluid tricritical point---the values of the nuclear masses at
which the intermediate changes of the phase diagram occur (the first
appearance of the tricritical point and the disappearance of the
liquid-gas critical point) cannot be obtained through simple
calculations.  Fortunately, this problem is free from the sign problem
and can be studied using quantum Monte-Carlo simulations~\cite{Youssef}.

Much of the discussion of this paper should be relevant in a more
general setting of van der Waals liquids, i.e., liquids consisting of
particles which interact with each other through a potential which has
a $r^{-6}$ power-law decay at long distances and a repulsive core at
short distances. It is possible that many features of the phase
diagrams would remain the same for other exponents of the power-law
tail of the potential.
It would be also interesting to examine the possible phase diagrams of
van der Waals quantum liquids in two dimensions.


The authors thank Massimo Boninsegni, Luca Delacr\'etaz,
Youssef Kora, Sergej Moroz, Ana Maria Rey,
Shiwei Zhang, and Wilhelm Zwerger for discussions and comments. This
work is supported by the U.S.\ Department of Energy, Office of
Science, Office of Nuclear Physics, within the framework of the Beam
Energy Scan Theory (BEST) Topical Collaboration and grant
No.\ DE-FG0201ER41195, by the U.S.\ DOE grant No.\ DE-FG02-13ER41958,
by a Simons Investigator grant and by the Simons Collaboration on
Ultra-Quantum Matter from the Simons Foundation.

\bibliography{UQHe}

\appendix

\section{Suppression of thermal non-tadpole diagrams}
\label{sec:supr-therm-non}

In our derivation of the three-dimensional effective field theory we have
evaluated only the tadpole diagrams.  We now show that non-tadpole graphs
are suppressed.  

Take as an example the correction to $g_{\rm eff}$
from the one-loop diagrams with two $g'$-vertices
\be
\Delta g=-\frac{g^{\prime2}}{\beta}\sum_{n\neq 0}\int_{\bm k}
  \bm k^4 \left[
  {1\over \left({2\pi n\over\beta}\right)^2+\left({\hbar^2\bm k^2\over 2m}-\mu\right)^2}+{4\over \left({-2\pi i n\over\beta}+{\hbar^2\bm k^2\over 2m}-\mu\right)^2} \right].
\ee
The $n$ summation can be done, but it is easy to see that the result is parametrically
\be
  \Delta g = \frac{g^{\prime2}\beta}{\lambda_T^7} F(-\beta\mu),
\ee
with some function $F$ that is finite when $(-\beta\mu)\to 0$.  
In the regime under consideration
\begin{equation}
  T\sim T_{\rm tri} \sim \frac{\hbar^2}m \left(\frac{|a|}{\sigma^4}\right)^{2/3},
  \qquad
  \lambda_T \sim \left(\frac{\sigma^4}{|a|} \right)^{1/3} ,
\end{equation}
the correction is much smaller than the bare value
\begin{equation}\label{Deltagg}
  \Delta g \sim \frac{\hbar^2}m \frac{|a|^{5/3}}{\sigma^{2/3}}
  \sim \left( \frac{|a|}\sigma \right)^{2/3} g \ll g .
\end{equation}


To proceed in a more general fashion, one can use the unit system with
$\hbar=m=1$, in which we can assign dimensions as in a nonrelativistic
theory
\begin{equation}
  [t] = -2,\quad [x] = [\lambda_T] = -1,\quad [\psi] = \frac32,\quad
  [g] = -1,\quad [g'] = -3, \quad  [G] = -4 .
\end{equation}
As integrating out modes with nonzero Matsubara frequency in the
regime $\beta\mu\ll1$ can bring out only powers of the thermal
wavelength $\lambda_T$, the contribution to a quantity $O$ of dimension
$\Delta_O$ from a loop diagram containing $N_g$, $N_{g'}$, and $N_G$
vertices of the respective types is of order
\begin{equation}\label{powercounting}
  \frac{g^{N_g} g^{\prime N_{g'}}G^{N_G}}
  {\lambda_T^{\Delta_O+N_g+3N_{g'}+4N_G}}
  \sim
  \left(\frac1\sigma\right)^{\Delta_O}
  \left(\frac{|a|}\sigma \right)^{\frac13\Delta_O + N_{g'}
     + \frac43 (N_g+N_G)}
\end{equation}
Thus, the leading loop contribution to a given vertex is from the diagrams
which minimize $N_{g'}+\frac43(N_g+N_G)$.  For example, the leading
contribution to $\mu_{\rm eff}$ comes from a diagram with $N_{g'}=1$, $N_g=N_G=0$, and the leading contribution to $g_{\rm eff}$ from a diagram with $N_G=1$,
$N_g=N_{g'}=0$.  These are tadpole diagrams.
The non-tadpole diagrams are supressed
compared to the leading tadpole diagrams by at least $(a/\sigma)^{2/3}$. 
For example, from Eq.~(\ref{powercounting})
we find that
the correction to $g_{\rm eff}$ coming
from a diagram with two $g'$ vertices ($N_{g'}=2$) is suppressed compared
by the tadpole diagram with $N_G=1$ by a factor of $(|a|/\sigma)^{2/3}$,
exactly as we have found in Eq.~(\ref{Deltagg}).

\section{Suppression of loops in the 3D effective field theory}
\label{sec:suppression-loops-3d}

So far we have treated the 3D effective field theory~(\ref{3Deff})
classically.  One can ask about the importance of loop corrections
within this 3D theory.  To answer  that question we use the unit system
$\hbar=m_{\rm eff}=1$ and rewrite the 3D effective theory in Eq.~(\ref{3Deff})
in the form (keeping also the higher-order term $g'$)
\begin{equation}\label{eq:L3D-m3D}
  \mathcal L_{\rm 3D}/\beta = \frac12 |\bm{\nabla}\psi_0|^2
  + \frac12 m^2_{\rm 3D}|\psi_0|^2
  + \frac{g_{\rm eff}}{2}|\psi_0|^4
  + \frac{g'}{2}(\bm\nabla |\psi_0|^2)^2
  + \frac{G}{6} |\psi_0|^6,
\end{equation}
where
\begin{equation}
  \label{eq:m3D}
  m_{\rm 3D}^2 = \mu_{\rm eff} \sim   \frac{a^2}{\sigma^4}\,.
\end{equation}
The dimensionless coupling constants controlling the loop corrections in
the 3D theory are given by 
\begin{equation}
  \label{eq:dimless-couplings}
    \frac{g_{\rm eff}T}{m_{\rm 3D}} \sim \left( \frac{|a|}\sigma\right)^{2/3}, \quad
  g'T m_{\rm 3D} \sim \left(\frac{|a|}\sigma\right)^{5/3} , \quad
  G T^2  \sim \left(\frac{|a|}\sigma \right)^{4/3} \,,
\end{equation}
and are all small in the regime $|a|/\sigma\ll1$ we consider.

One can also ask if the higher-derivative or higher-order terms
contribute to the surface tension calculated in Section~\ref{sec:first-order-phase}. To
answer that question
 we need to count powers of the field as $|\psi_0|^2=n_{\rm cond}\sim
a/\sigma^4$ and each gradient as $\bm\nabla\sim 1/\xi\sim a/\sigma^2$. Thus, for
example, the contribution of the derivative interaction term in Eq.~(\ref{eq:L3D-m3D}) can be estimated as
\begin{equation}
  \label{eq:1}
  g'(\bm\nabla|\psi_0|^2)^2\sim\frac{1}{\sigma^5}\,\left(\frac{|a|}{\sigma}\right)^4\,,
\end{equation}
which is suppressed compared to the contribution of all other terms in
Eq.~(\ref{eq:L3D-m3D}) such as, e.g.,
$G|\psi_0|^6\sim (1/\sigma^5)(|a|/\sigma)^3$.

\end{document}